\documentclass[prl,preprint,groupedaddress,showpacs]{revtex4-1}

\usepackage{amsmath}
\usepackage{amssymb}
\usepackage{amsfonts}
\usepackage{mathrsfs}
\usepackage{epsfig}
\usepackage{bm}
\usepackage{multirow}

\begin{document}

\title{Amplification and generation of ultra-intense twisted laser pulses via stimulated Raman scattering}
\author{J. Vieira$^{1,*}$,R.M.G.M. Trines$^2$, E.P. Alves$^{1}$, R.A. Fonseca$^{1,3}$,J.T.Mendon\c{c}a$^1$,R. Bingham$^{2}$, P.Norreys$^{2,4}$, L.O.Silva$^{1}$}
\affiliation{$^1$GoLP/Instituto de Plasmas e Fus\~{a}o Nuclear,  Instituto Superior T\'{e}cnico, Universidade de Lisboa, Lisbon, Portugal}
\affiliation{$^2$Central Laser Facility, STFC Rutherford Appleton Laboratory, Didcot, OX11 0QX, United Kingdom}
\affiliation{$^3$DCTI/ISCTE Lisbon University Institute, 1649-026 Lisbon, Portugal}
\affiliation{$^4$Department of Physics, University of Oxford, Oxford OX1 3PU, UK}
\address{$^*$Corresponding author: jorge.vieira@ist.utl.pt}
\today

\maketitle

\textbf{Twisted Laguerre-Gaussian lasers, with orbital angular momentum and characterised by doughnut shaped intensity profiles, provide a transformative set of tools and research directions in a growing range of fields and applications, from super-resolution microcopy and ultra-fast optical communications to quantum computing and astrophysics. The impact of twisted light is widening as recent numerical calculations provided solutions to long-standing challenges in plasma-based acceleration by allowing for high gradient positron acceleration. The production of ultrahigh intensity twisted laser pulses could then also have a broad influence on relativistic laser-matter interactions. 
Here we show theoretically and with \textit{ab-initio} three-dimensional particle-in-cell simulations, that stimulated Raman backscattering can generate and amplify twisted lasers to Petawatt intensities in plasmas. This work may open new research directions in non-linear optics and high energy density science, compact plasma based accelerators and light sources.}

\section{Introduction}

The seminal work by L. Allen~\emph{et al}~\cite{bib:allen_pra_1992} on lasers with orbital angular momentum (OAM) has initiated a path of significant scientific developments that can potentially offer new technologies in a growing range of fields, including microscopy~\cite{bib:jesacher_prl_2005} and imaging~\cite{bib:jack_prl_2009}, atomic~\cite{bib:andersen_prl_2006} and nano-particle manipulation~\cite{bib:padgett_review_2011}, ultra-fast optical communications~\cite{bib:wang_nphoton_2013,bib:bozinovic_science_2013}, quantum computing~\cite{bib:terriza_progress_2007} and astrophysics~\cite{bib:tamburini_nphys_2011}. At intensities beyond material breakdown thresholds, it has been recently shown through theory and simulations that intense (with $\gtrsim 10^{18}~\mathrm{W cm^{-2}}$ intensities) and short (with $10-100~\mathrm{fs}$ durations) twisted laser beams could also excite strongly non-linear plasma waves suitable for high gradient positron acceleration in plasma accelerators~\cite{bib:vieira_prl_2014}. As a result of their importance, many techniques have emerged to produce Laguerre-Gaussian lasers over a wide range of frequencies~\cite{bib:padgett_phystoday_2004}. Common schemes employ spiral phase plates or computer generated holograms to generate visible light with OAM, non-linear optical media for high-harmonic generation and emission of XUV OAM lasers~\cite{bib:corkum_prl_2014,bib:shao_pra_2013} or spiral electron beams in free-electron lasers to produce OAM x-rays~\cite{bib:hemsing_prl_2012,bib:hemsing_natphys_2013}.

Optical elements such as spiral phase plates are designed for the production of laser beams with pre-defined OAM mode contents. Novel and more flexible mechanisms capable of producing and amplifying beams that with arbitrary, well defined orbital angular momentum states, using a single optical component, would then be interesting from a fundamental point of view, while also benefiting experiments where OAM light is relevant. In addition, the possibility of extending these mechanisms to the production and amplification of laser pulses with relativistic intensities, well above the damage thresholds of optical devices, could also open exciting perspectives for high energy density science and applications. The use of a plasma as the optical medium is a potential route towards the production of OAM light with relativistic intensities~\cite{bib:gordon_ol_2009}. Although other routes may be used to produce high intensity OAM laser pulses, for instance by placing spiral phase plates either at the start or at the end of a laser amplification chain~\cite{bib:oam_plasma,bib:brabetz_pop_2015}, the use of plasmas can potentially lead to the amplification of OAM light to very high powers and intensities. Plasmas also allow for greater flexibility in the level of orbital angular momentum in the output laser beam than other more conventional techniques. 

Here, we show that stimulated Raman scattering processes in non-linear optical media with a Kerr non-linearity can be used to generate and to amplify OAM light. Plasmas, optical fibres and non-linear optical crystals are examples of non-linear optical media with Kerr non-linearity. Although optical parametric oscillators have also been used to transfer OAM from a pump to down converted beams~\cite{bib:martinelli_pra_2004}, here we explore the creation of new OAM states absent from the initial configuration, according to simple selection rules. We also demonstrate that stimulated Raman scattering processes can generate and amplify OAM light even in scenarios where no net OAM is initially present. To this end, we use an analytical theory, valid for arbitrary transverse laser field envelope profiles complemented by the first three-dimensional \emph{ab initio} particle-in-cell simulation of the process using the PIC code Osiris~\cite{bib:osiris}, considering that the optical medium is a plasma. Starting from recent experimental and theoretical advances~\cite{bib:ren_nphys_2007,bib:trines_nphys_2011,bib:trines_prl_2011} our simulations and theoretical developments show that stimulated Raman processes could pave the way to generate OAM light in non-linear optical media and that the non-linear optics of plasmas~\cite{bib:michel_prl_2014,bib:mori_ieee_1997} could provide a path to generate and amplify OAM light to relativistic intensities~\cite{bib:shvets_prl_1998,bib:malkin_prl_1999,forslund_pf_1975,bib:malkin_pop_2000}.

\section{Results}

\subsection{Theoretical model}

We illustrate our findings considering that the non-linear optical medium is a plasma. Extension to other materials is straightforward. In a plasma, stimulated Raman backscattering is a three-wave mode coupling mechanism in which a pump pulse (frequency $\omega_0$ and wavenumber $k_0$), decays into an electrostatic, or Langmuir, plasma wave (frequency $\omega_{\mathrm{p}}$ and wavenumber $2 k_0-\omega_{\mathrm{p}}/c$) and into a counter-propagating seed laser (frequency $\omega_1 = \omega_0-\omega_{\mathrm{p}}$ and wavenumber $k_1 = \omega_{\mathrm{p}}/c-k_0$). The presence of OAM in the pump and/or seed results in additional matching conditions that ensure the conservation of the angular momentum carried by the pump when the pump itself decays into a scattered EM wave and a Langmuir wave~\cite{bib:medonca_prl_2009}. These additional matching conditions, which 
are explored in more detail in the Supplementary Information in Supplementary Notes 1-4 and Supplementary Figs. 1-3), 
correspond to selection rules for the angular momentum carried by each laser and plasma wave. Here, we illustrate key properties of OAM generation and amplification, by exploring different seed and pump configurations.

In order to derive a model capable of predicting stimulated Raman scattering OAM selection rules, we start with the general equations describing Stimulated Raman scattering given by $D_0 \mathbf{A}_0 = \omega_{\mathrm{p}}^2 \delta n \mathbf{A}_1$, $D_1 \mathbf{A}_1 = -\omega_{\mathrm{p}}^2 \delta n^{*} \mathbf{A}_0$ and $D_p \delta n= e^2 k_{\mathrm{p}}^2/(2 m_e^2) \mathbf{A}_0\cdot \mathbf{A}_1$, where $D_{0,1}= c^2 \left(\nabla_{\perp}^2+2 i k_{0,1}\partial_z\right)+ 2 i \omega_{0,1}\partial t$, $D_p = 2 i \omega_{\mathrm{p}}\partial t$. Moreover, $\mathbf{A}_{0,1}$ is the envelope of the pump/seed laser, with complex amplitude given by $\mathbf{A}_1(t,\mathbf{r}_{\perp}) \exp\left[i k_{0,1} z- i \omega_{0,1} t \right] + c.c.$, where $t$ is the time and $z$ the propagation distance. We note that $\mathbf{A}_{0,1}$ are arbitrary functions of the transverse coordinate $\mathbf{r}_{\perp}$. The complex amplitude of the plasma density perturbations is $\delta n(t,\mathbf{r}_{\perp}) \exp\left[i k_{\mathrm{p}} z- i \omega_{\mathrm{p}} t \right] + c.c.$, where $k_{\mathrm{p}} = \omega_{\mathrm{p}}/c$ is the plasma wavenumber, $\omega_{\mathrm{p}} = \sqrt{e^2n_0/\epsilon_0 m_e}$ the plasma frequency, $m_e$ the mass of the electron, $\epsilon_0$ the vacuum electric permittivity, and $e$ the elementary charge. Although these general equations can be used to retrieve the selection rules for the orbital angular momentum that will be explored throughout this paper, it is possible to derive exact solutions in the long pulse limit, where $k_{0,1}\partial_z \ll \omega_{0,1} \partial_t$ and in the limit where the pump laser contains much more energy than the seed laser energy. In this case, since $\partial_t \mathbf{A}_0^2 \sim \partial_t \mathbf{A}_1^2$, and $\mathbf{A}_0\gg \mathbf{A}_1$ (pump has more energy than seed) then $\partial_t \mathbf{A}_0 \ll \partial_t \mathbf{A}_1$ (this condition is strictly satisfied in our simulations when new modes are created and until their energy becomes comparable to the energy in the pump pulse). In this case, it is possible to show that stimulated Raman scattering of a seed beam $\mathbf{A}_1$ from a pump beam $\mathbf{A}_0$ creates a plasma wave density perturbation given by:
\begin{equation}
\label{eq:deltan}
\delta n^*(t,\mathbf{r}_{\perp}) =  i \frac{e^2 k_{\mathrm{p}}^2}{4 \omega_{\mathrm{p}} m_e^2} \left[\mathbf{A}_{0}^* \cdot \mathbf{A}_{1}(t=0)\right] \frac{\sinh\left(\Gamma t \right)}{\Gamma},
\end{equation}
\begin{equation}
\label{eq:growthrate}
\Gamma^2(\mathbf{r}_{\perp}) = \frac{e^2 k_{\mathrm{p}}^2 \omega_{\mathrm{p}}^2}{8 \omega_{\mathrm{p}} \omega_{1}m_e^2} |\mathbf{A}_{0}|^2, 
\end{equation}
where $\Gamma$ is the growth rate at which the plasma amplitude grows as the interaction progresses, and where $\mathbf{r}_{\perp}$ is the transverse position. The amplification of the seed is given by:
\begin{equation}
\label{eq:amplification}
\mathbf{A}_{1}(t,\mathbf{r}_{\perp}) = \left(\mathbf{A}_{1}(t=0)\cdot \frac{\mathbf{A}_{0}^*}{|\mathbf{A}_{0}|}\right) \frac{\mathbf{A}_{0}}{|\mathbf{A}_{0}|} \cosh\left(\Gamma t\right) + C, \\
\end{equation}
where $(\omega_1,k_1)$ are the frequency and wavenumber of the seed laser pulse, %
and where $C$ is a constant of integration. %
The derivation of Eqs.~(\ref{eq:deltan}), (\ref{eq:growthrate}) and (\ref{eq:amplification}), presented in detail in the Supplementary Note 1, assumes that the pump and the seed satisfy the frequency matching conditions stated above, being valid for arbitrary transverse laser envelope profiles as long as the paraxial equation is satisfied. Neglecting pump depletion does not change the selection rules for the orbital angular momentum discussed for the remainder of this work. Unless explicitly stated, the generic expression for the pump vector potential (or electric field) is $\mathbf{A}_0 = A_{0\mathrm{x}} \exp(i \ell_{0\mathrm{x}} \phi) \mathbf{e}_\mathrm{x} + A_{0\mathrm{y}} \exp(i \ell_{0\mathrm{y}} \phi) \mathbf{e}_\mathrm{y}$, where $(\mathbf{e}_\mathrm{x},\mathbf{e}_\mathrm{y})$ are the unit vectors in the transverse x and y directions and $\phi$ the azimutal angle. Similarly, the generic expression for the seed vector potential is $\mathbf{A}_1 = A_{1\mathrm{x}} \exp(i \ell_{1\mathrm{x}} \phi) \mathbf{e}_{\mathrm{x}} + A_{1\mathrm{y}} \exp(i \ell_{1\mathrm{y}} \phi) \mathbf{e}_{\mathrm{y}}$. Selection rules can then be generally derived by inserting these expressions into the factor $
\left(\mathbf{A}_{1}(t=0)\cdot \mathbf{A}_{0}\right)\mathbf{A}_{0}$ in Eq.~(\ref{eq:amplification}). Although we have assumed that the plasma is the optical medium, other nonlinear optical media with Kerr non-linearity will also exhibit similar phenomena. 

\subsection{Particle-in-cell simulations}

\begin{figure}
\centering\includegraphics[width=\columnwidth]{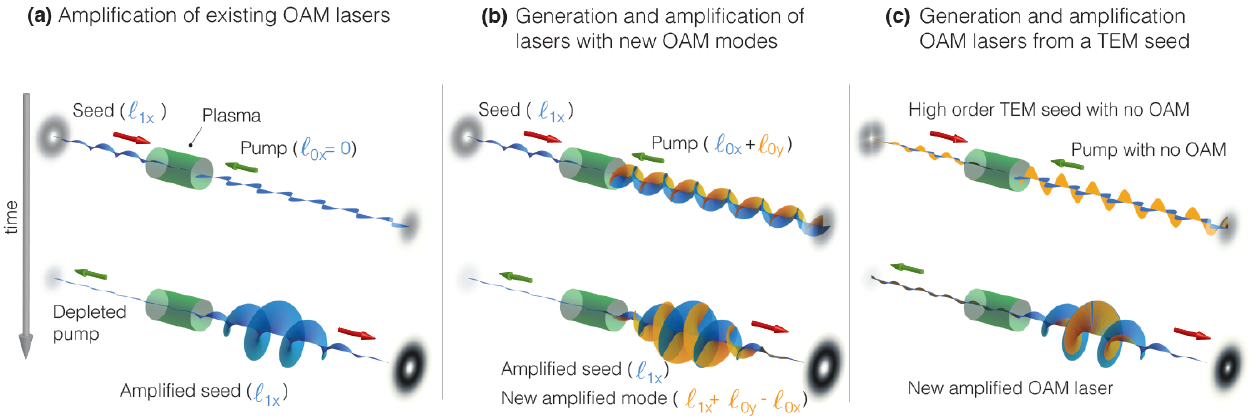}
\caption{Generation and amplification of orbital angular momentum (OAM) lasers via stimulated Raman Backscattering. In Raman amplification, a long pump laser transfers its energy to a short, counter-propagating seed laser in a plasma. The process depletes the pump laser pulse energy and enhances the intensity of the seed laser. The seed/pump lasers propagate in the direction of the red/green arrow, respectively. Polarisation in the x/y direction is represented by blue/orange lasers respectively. The position of the plasma, relative to the seed and pump lasers, is shown by the green cylinders. The back/front projections show the intensity profile of the closest laser. Panel (a) shows a setup leading to the amplification of a seed with OAM. Panel (b) shows the generation and amplification of a new OAM modes. Panel (c) shows the generation and amplification of a new OAM laser in a configuration with no initial OAM.}
\label{fig:setup}
\end{figure}

We will now use Eq.~(\ref{eq:amplification}) to explore OAM generation and amplification in three separate classes of initial set-ups, all identified in Fig.~\ref{fig:setup}. We start by studying the case of the amplification of existing OAM modes. 
Figure~\ref{fig:setup}(a) illustrates the process in a setup leading to the amplification of a seed in an arbitrary, single state of orbital angular momentum $\ell_1$ in a plasma using a counter-propagating Gaussian pump laser without OAM, but the mechanism is trivially generalised for a pump with arbitrary OAM $\ell_0$.  
We can then assume a pump linearly polarised in the x direction with OAM $\ell_{0\mathrm{x}}$, which decays into a Langmuir plasma wave with OAM $\ell_{\mathrm{p}} = \ell_{0\mathrm{x}} - \ell_{1\mathrm{x}}$ and into a seed, also linearly polarised in the x direction with OAM $\ell_{1\mathrm{x}}$. Making these substitutions into Eq.~(\ref{eq:amplification}) confirms that the amplification of $A_1$ retains the initial seed OAM. For the specific example in Fig.~\ref{fig:setup}(a) where $\ell_{0\mathrm{x}}=0$, direct substitution of $\mathbf{A}_{0}\sim \exp(i \ell_{0\mathrm{x}} \phi) \mathbf{e}_{\mathrm{x}}$ and of $\mathbf{A}_1 \sim \exp(i \ell_{1\mathrm{x}}\phi) \mathbf{e}_{\mathrm{x}}$ in Eq.~(\ref{eq:deltan}) shows that the plasma wave density perturbations $\delta n \sim \exp[i (\ell_{0\mathrm{x}}-\ell_{1\mathrm{x}})\phi]$ have OAM $\ell_{\mathrm{p}} = \ell_{0\mathrm{x}}-\ell_{1\mathrm{x}}=-\ell_{1\mathrm{x}}$, i.e. the plasma wave absorbs the excess OAM that may exist between the pump and the seed 
(see Supplementary Note 2 and Supplementary Fig. 1 for several examples illustrating angular and linear momentum matching conditions demonstrating that the plasma wave always absorbs the excess OAM between pump and seed.) 
The scheme is thus ideally suited to amplify an existing OAM seed using a long Gaussian pump without OAM. The amplification of circularly polarised OAM lasers, with both spin and orbital angular momentum, obeys similar selection rules. For amplification to occur in this case, and like stimulated Raman backscattering of circularly polarised Gaussian lasers, both seed and pump need to be polarised with the same handedness either in $\mathbf{e}_{+}=\mathbf{e}_{\mathrm{x}}+i \mathbf{e}_{\mathrm{y}}$ or in $\mathbf{e}_{-}=\mathbf{e}_{\mathrm{x}}-i \mathbf{e}_{\mathrm{y}}$.

\begin{figure}
\centering\includegraphics[width=\columnwidth]{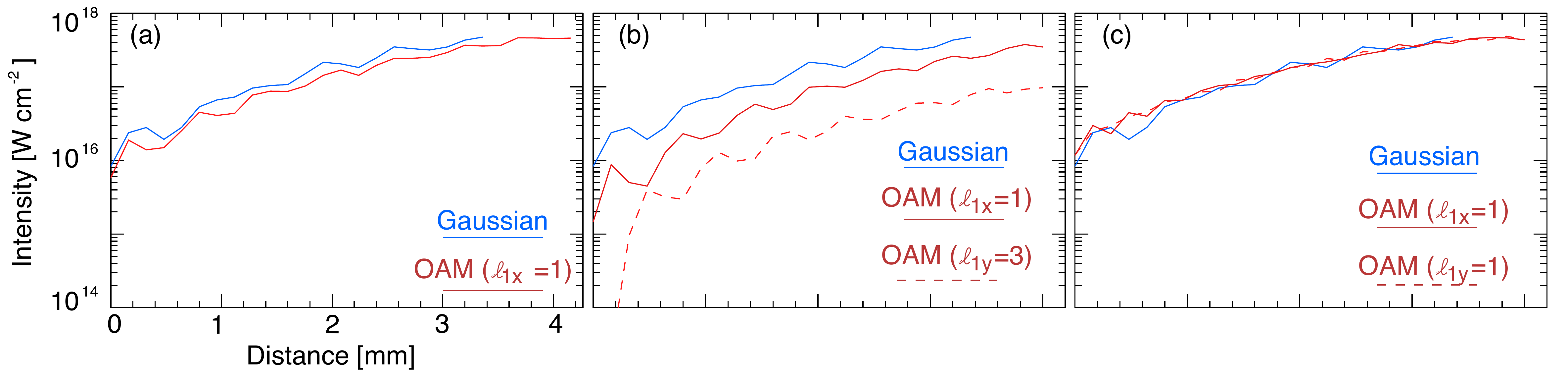}
\caption{Simulation results shown the evolution of seed laser intensities (I) as a function of the propagation distance (z) for the different configurations of Fig.~\ref{fig:setup}. Blue refers to the amplification of a Gaussian seed by a Gaussian pump. The initial laser configuration in each panel (a)-(c) corresponds to the initial setup illustrated by each corresponding panel (a)-(c) in Fig.~\ref{fig:setup}. Panel (a) shows the amplification of a seed with $\ell_{1\mathrm{x}}=1$ using a long Gaussian pump (red). Panel (b) shows the generation and amplification of a new OAM mode with $\ell_{1\mathrm{y}}=3$ (dashed-red) and of an existing mode with $\ell_{1\mathrm{x}}=1$ (solid-red) from an orbital angular momentum (OAM) pump polarised in two directions with $\ell_{0\mathrm{y}}=2$ and $\ell_{0\mathrm{x}}=0$. Panel (c) shows the amplification of a new OAM mode with $\ell_1=1$ from a TEM seed with no net initial OAM and from a Gaussian pump. See Table 1 for simulation parameters.
}
\label{fig:amp-rate}
\end{figure}

Figure~\ref{fig:amp-rate}a illustrates 3D simulation results showing the amplification of an $\ell_1=1$, linearly polarised seed from a linearly polarised Gaussian pump. Simulation parameters are stated in Table 1. Figure~\ref{fig:amp-rate}a shows that the growth rate for the amplification process is nearly indistinguishable from stimulated Raman amplification of Gaussian lasers. 
In agreement with Eq.~(\ref{eq:growthrate}), this result also indicates that, in general, the overall amplification process is OAM independent.

Stimulated Raman scattering also provides a mechanism to create new OAM modes (i.e. modes that are absent from the initial pump/seed lasers) and amplify them to very high intensities. 
Figure~\ref{fig:setup}(b), 
illustrates the process schematically. The pump electric fields can have different OAM components in both transverse directions x and y. Each component is represented in blue and in orange in Fig.~\ref{fig:setup}(b). The pump electric field component in x has OAM $\ell_{0\mathrm{x}}$. The pump electric field component in y has OAM $\ell_{0\mathrm{y}}$. The initial seed electric field contains an OAM $\ell_{1\mathrm{x}}$ component in the x direction.  After interacting in the plasma, the pump becomes depleted and a new electric field component appears in the seed with OAM given by $\ell_{1\mathrm{y}}=\ell_{1\mathrm{x}}+\ell_{0\mathrm{y}}-\ell_{0\mathrm{x}}$.

The process can be physically understood by examining the couplings between the plasma and light waves in the example considered above. Initially, a plasma wave will be excited due to beating pump and seed components that have their electric fields pointing in the transverse x direction. According to Eq.~(\ref{eq:deltan}), the plasma wave OAM is $\ell_{\mathrm{p}} = \ell_{0\mathrm{x}}-\ell_{1\mathrm{x}}$. This plasma wave ensures OAM conservation for the pump and seed electric field components in the x direction. The (same) plasma wave also couples the pump and seed modes with electric field components pointing in the y direction. Thus, $\ell_{\mathrm{p}} = \ell_{1\mathrm{y}}-\ell_{0\mathrm{y}}$ must also hold in order to ensure conservation of angular momentum. This implies the generation of a new seed component with electric field polarised in y so that OAM is conserved at all times and in both components. The OAM of the new seed component is thus $\ell_{1\mathrm{y}} = \ell_{\mathrm{p}} +\ell_{0\mathrm{y}} = \ell_{1\mathrm{x}}-\ell_{0\mathrm{x}}+\ell_{0\mathrm{y}}$. 

Alternatively, this selection rule can also be found by examining Eq.~(\ref{eq:amplification}). Direct substitution of a pump profile with $\mathbf{A}_0\sim \exp\left(i \ell_{0\mathrm{x}} \phi\right) \mathbf{e}_{\mathrm{x}} + \exp\left(i \ell_{0\mathrm{y}} \phi\right) \mathbf{e}_{\mathrm{y}}$ and of an initial seed profile with $\mathbf{A}_1 \sim \exp\left(i \ell_{1\mathrm{x}} \phi\right) \mathbf{e}_{\mathrm{x}}$ then leads to the generation of a new seed with $\exp\left[i (\ell_{1\mathrm{x}}+\ell_{0\mathrm{y}}-\ell_{0\mathrm{x}})\phi \right] \mathbf{e}_{\mathrm{y}}$. The same selection rules would also hold if the pump consists of combination of a right and left handed circularly polarised modes, each with different OAM, and the seed initially contains only a left or right handed circularly polarised mode. In this case a new seed component would appear with right or left handed circular polarisation. The new mode is created to ensure conservation of orbital angular momentum. The selection rules are identical as long as polarisations in $\mathbf{e}_{\mathrm{x}}$ / $\mathbf{e}_{\mathrm{y}}$ are replaced by polarisations in $\mathbf{e}_{+}$ / $\mathbf{e}_{-}$ (where $\mathbf{e}_{\pm}=\mathbf{e}_{x}\pm i \mathbf{e}_{y}$). This setup provides a robust mechanism for the production and amplification of a new and well defined OAM mode, absent from the initial set of lasers. 
The generation of a new OAM mode when a linearly polarised seed interacts with a pump with electric field components in the two orthogonal directions is also illustrated in Supplementary Note 3 and Supplementary Fig. 2.

Figure~\ref{fig:new-oam} shows a result of a 3D PIC simulation illustrating the production of a new seed mode with $\ell_{1\mathrm{y}}=3$, which is initially absent from the simulation, from a pump with $\ell_{0\mathrm{y}}=2$, $\ell_{0\mathrm{x}} = 0$ and an initial seed with $\ell_{1\mathrm{x}}=1$ (simulation parameters given in Table 1). 
Figure~\ref{fig:new-oam} presents several distinct signatures of the new OAM mode with $\ell_{1\mathrm{y}}=3$. The laser vector potential shows helical structures, which indicate that the new mode has OAM. The normalised vector potential forms a pattern that repeats each 3 turns, which turn in the clock-wise direction from the front to the back of the pulse, a signature for $\ell_{1\mathrm{y}}=3$. Field projections in the (x,y) plane also show a similar pattern further confirming that the new OAM mode has $|\ell_{1\mathrm{y}}|=3$. The change in color from blue-green in (a) to (green-red) in (b) is a clear signature for the intensity amplification of the new seed mode. Intensity was calculated using $I \left[\mathrm{W/cm^2}\right]=1.27 \times 10^{18} a_0^2/\lambda_0^2 \left[\mathrm{\mu m}\right]$, where $\lambda_0=0.8\mu m$ is the central laser wavelength.
Figure~\ref{fig:new-oam} also shows that the amplified laser envelope acquires a bow shaped profile~\cite{bib:trines_nphys_2011,bib:trines_prl_2011,bib:fraiman_pop_2002}, a key signature of Raman amplification identified in~\cite{bib:trines_nphys_2011}. In agreement with theory (see Eqs.~(\ref{eq:growthrate}) and (\ref{eq:amplification})), Fig.~\ref{fig:amp-rate}b shows that the new $\ell_{1\mathrm{y}}=3$ mode and the existing $\ell_{1\mathrm{x}}=1$ OAM mode amplify at nearly coincident growth rates. Still, since it grows from initially higher intensities, the existing $\ell_{1\mathrm{x}}=1$ mode reaches higher final intensities than the new OAM mode with $\ell_{1\mathrm{y}}=3$. The generation of the new modes in Fig.~\ref{fig:amp-rate}b also illustrates the transition from the regime where the depletion of the pump is negligible (small signal, exponential growth) to the regime where the depletion of the pump is not negligible (strong signal, linear growth). Hence, Fig.~\ref{fig:amp-rate}b shows an exponential growth of the new mode up to $z<0.6~\mathrm{mm}$). For $z>0.6~\mathrm{mm}$, the energy in the new seed becomes comparable to the energy contained in the pump pulse. As a result, the growth slows down significantly, becoming linear with the propagation distance~\cite{bib:malkin_prl_1999}.

L. Allen~\emph{et al}~\cite{bib:allen_pra_1992} showed that particular superpositions of Hermite-Gaussian modes 
(also called Transverse Electro Magnetic or TEM modes)
are mathematically equivalent to Laguerre-Gaussian modes. %
Since the transverse amplitude distribution of high order (transverse) laser modes is usually described by a product of Hermite-Gaussian polynomials, which is also usually associated with TEM modes, this result paved the way for experimental realisation of vortex light beams with orbital angular momentum from existing TEM laser modes. It is thus interesting and important to explore if and how stimulated Raman backscattering can be used to generate and amplify light with orbital angular momentum from Hermite-Gaussian laser beams, i.e. from initial configurations with no net OAM. Figure~\ref{fig:setup}c illustrates the process. From now on, we refer to each Hermite-Gaussian beam as a TEM$_{m,n}$ laser where $(m,n)$ represents the Hermite-Gaussian mode. The TEM mode electric field is given by Eq.~(\ref{eq:hermite}) (see Methods Section). %
We consider first a Gaussian pump linearly polarised at $45^{\circ}$, i.e. having similar electric field amplitudes in both transverse directions x and y. The Gaussian pump, can then also be written as $\mathbf{A}_0 \sim \mathrm{TEM}_{00}\left( \mathbf{e}_{\mathrm{x}}+\mathbf{e}_{\mathrm{y}} \right)$. In addition, we assume a seed with a $\mathrm{TEM}_{10}$ mode electric field component in x and with a $\mathrm{TEM}_{01}$ mode electric field component in y. The two seed modes are $\pi/2$ out of phase with respect to each other. The seed is given by $\mathbf{A}_1 \sim \mathrm{TEM}_{10} \mathbf{e}_{\mathrm{x}} + i \mathrm{TEM}_{01} \mathbf{e}_{\mathrm{y}}$. This setup is represented in Fig.~\ref{fig:setup}(c), where blue and orange colours in  refer to the pump and seed components polarised in the x and in the y directions respectively.

\begin{figure}
\centering\includegraphics[width=\columnwidth]{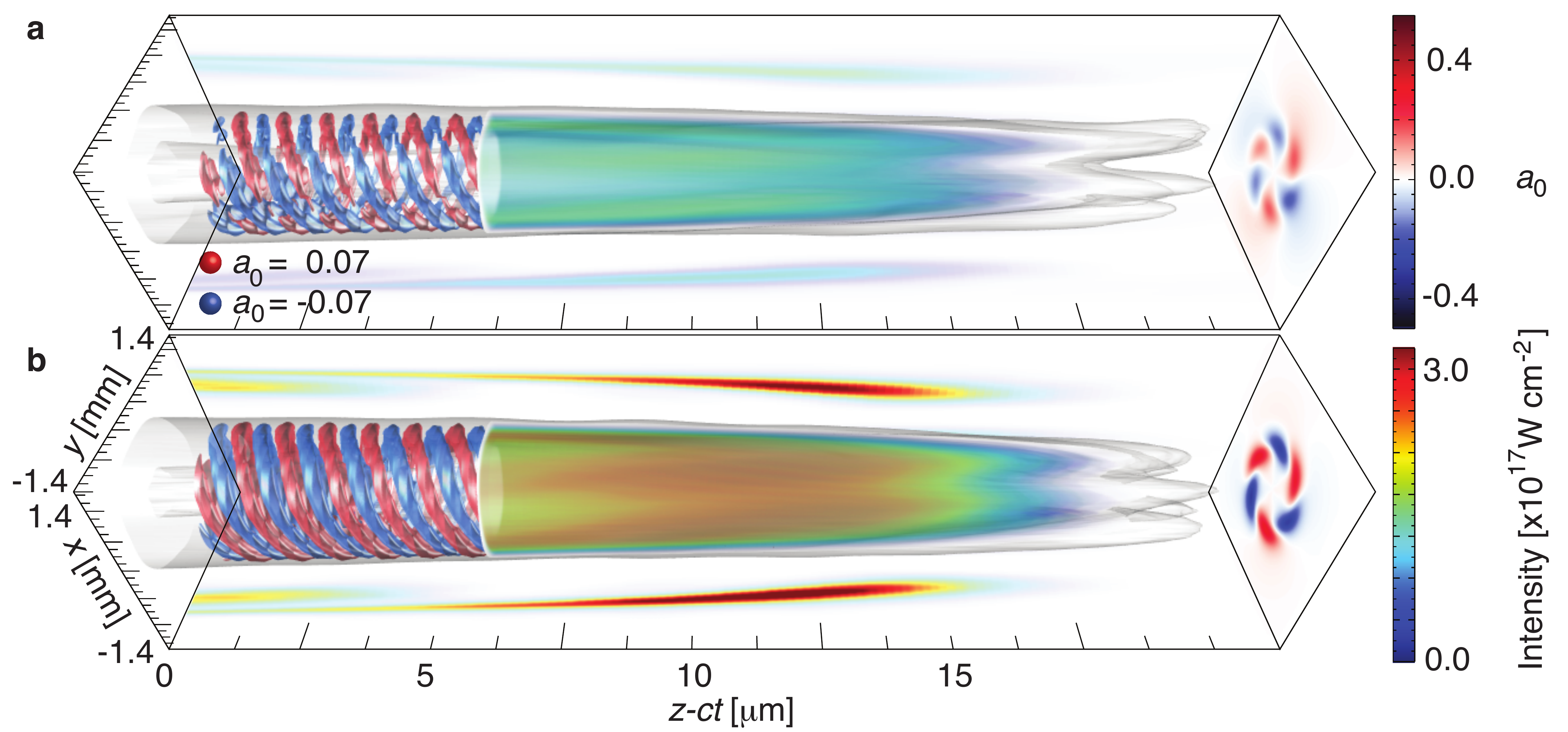}
\caption{Simulation results showing the generation and amplification of a new orbital angular momentum (OAM) modes. The new modes with $\ell_{1\mathrm{y}}=3$ grow from a seed with $\ell_{1\mathrm{x}}=1$ and a linearly polarised pump with a Gaussian profile in the x direction and with an OAM $\ell_{0\mathrm{y}}=2$ in the y direction. Panel (a) refers to $z=2~\mathrm{mm}$, and panel (b) to $z=6.22~\mathrm{mm}$. The initial setup is illustrated in Fig.~\ref{fig:setup}b. Projections in the (x,z) and (y,z) planes show intensity profile slices at the mid-plane of the OAM mode (blue-green-red colours). Projections in the (x,y) plane (blue-white-red) show the normalised vector potential ($a_0$) field envelope of the new OAM mode at the longitudinal slice where the laser intensity is maximum. The envelope of the 3D laser intensity is also shown for $z > 6.25$ mm in blue-green-red colours, and normalised vector potential isosurfaces for $z<6.25$ mm in blue and in red. 
}
\label{fig:new-oam}
\end{figure}

Although this setup has no initial OAM, since both pump and seed have no OAM, it results in the generation and amplification of an OAM mode with $\ell_1 = 1$. In order to understand the OAM generation mechanism we first consider Eq.~(\ref{eq:deltan}). According to Eq.~(\ref{eq:deltan}), the beating between the $\mathrm{TEM}_{10}$ seed with the Gaussian pump in the x direction will drive a $\mathrm{TEM}_{10}$ daughter plasma wave component. The beating between the $i \mathrm{TEM}_{01}$ seed and Gaussian pump in the y direction will drive a $i \mathrm{TEM}_{01}$ daughter plasma wave component. These two plasma wave components are $\pi/2$ out of phase with respect to each other. Hence, the resulting plasma wave will be a combination of TEM modes given by $\delta n \sim \mathrm{TEM}_{10} - i\mathrm{TEM}_{01}$, where the $i$ denotes the phase difference between modes. According to Ref.~\cite{bib:allen_pra_1992}, this mode combination is equivalent to a Laguerre-Gaussian mode with $\ell_{\mathrm{p}}=-1$. In order to conserve angular momentum, a new seed component with $\ell_1=1$ will then have to be generated and amplified in the direction of polarisation of the pump. 
This process is also illustrated in Supplementary Note 4 and Supplementary Fig. 3.

It is also possible to reach this conclusion by substituting in Eq.~(\ref{eq:amplification}) the expressions for initial Gaussian pump transverse profile [$\mathbf{A}_{0} \sim \mathrm{TEM}_{00}\left(\mathbf{e}_{\mathrm{x}}+\mathbf{e}_{\mathrm{y}}\right)$] and initial TEM seed profile ($\mathbf{A}_1 \sim \mathrm{TEM}_{10}\mathbf{e}_{\mathrm{x}}+i \mathrm{TEM}_{01}\mathbf{e}_{\mathrm{y}}$). This substitution yields a new seed transverse profile given by $\mathbf{A}_{1} \sim \left(\mathrm{TEM}_{10} + i \mathrm{TEM}_{01}\right) \left(\mathbf{e}_{\mathrm{x}}+\mathbf{e}_{\mathrm{y}}\right)$, corresponding to a Laguerre-Gaussian mode with $\ell_1 = 1$~\cite{bib:allen_pra_1992}. 
Similarly, a circularly polarised Gaussian pump and a $\mathbf{A}_1\sim \mathrm{TEM}_{10} \mathbf{e}_{\mathrm{x}}+\mathrm{TEM}_{01} \mathbf{e}_{\mathrm{y}}$ seed (i.e. without phase difference between the $\mathrm{TEM}_{10}$ and $\mathrm{TEM}_{01}$ modes) would also lead to a new seed component with $|\ell_1|=1$. The plasma can then be viewed as a high intensity mode converter. %

\begin{figure}
\centering\includegraphics[width=\columnwidth]{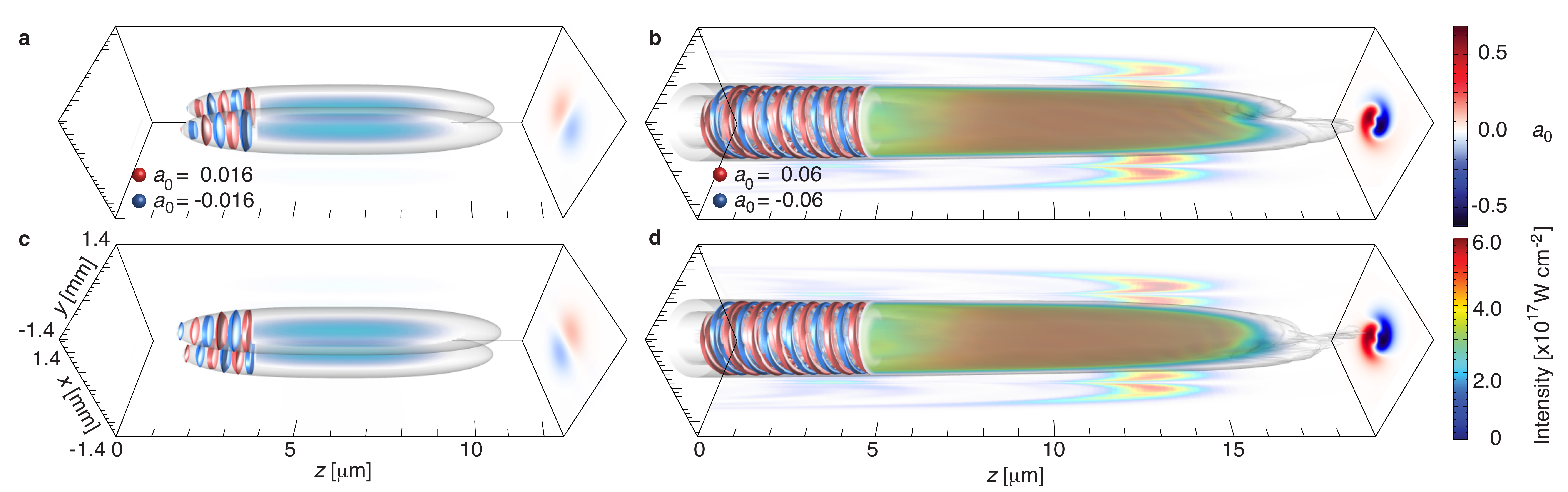}
\caption{Simulation result showing the generation and amplification of a new orbital angular momentum (OAM) mode from initial configurations with no net OAM. The new mode is linearly polarised in x and in y with $\ell_{1\mathrm{x}}=\ell_{1\mathrm{y}}=-1$ from an initial seed polarised in the x direction with a $\mathrm{TEM}_{01}$ mode and in the y direction with a $\mathrm{TEM}_{10}$ mode that is $\pi/2$ out of phase with respect to the $\mathrm{TEM}_{01}$ mode polarised in x. The pump is a Gaussian laser linearly polarised at $45^{\circ}$. The initial laser setup corresponds to Fig.~\ref{fig:setup}c. The meaning of the color scales and physical quantities plotted in all panels are identical to Fig.~\ref{fig:new-oam}. The values of the laser vector potential illustrated by the iso-surfaces of panel (a) and (c) are shown by the spheres in panel (a). Those iso-surface values for panels (b) and (d) are indicated in panel (b). Panels (a) and (c) show the initial seed TEM modes in x and y directions respectively. Panels (b) and (d) show the new OAM mode electric field components at $z=3.5~\mathrm{mm}$ in  $x$ (b) and in $y$ (d).}
\label{fig:tem-oam}
\end{figure}

Figure~\ref{fig:tem-oam} shows results from a 3D simulation that confirms these predictions (see Table 1 for simulation parameters). 
The simulation setup follows the example of Fig.~\ref{fig:setup}c described earlier. Simulations show that stimulated Raman scattering leads to a new OAM mode with $\ell_1=1$ linearly polarised at $45^{\circ}$. Figure~\ref{fig:amp-rate}c shows that the amplification rates are comparable to the other typical scenarios shown in Fig.~\ref{fig:amp-rate}a-b. The change on the field topology of the seed normalised vector potential shown in Fig.~\ref{fig:tem-oam}, from plane isosurfaces to helical isosurfaces, indicates the generation of a laser with orbital angular momentum from a configuration with no net OAM. Normalised vector potential iso-surfaces, and projection in the yz direction, form a pattern that repeats each turn and that rotates clock-wise from the front to the back of the pulse, indicating an OAM with $\ell_{1\mathrm{y}}=1$.

\section{Discussion}

We have so far assumed that the lasers are perfectly aligned. In experiments, however, the beams can only be aligned within a certain precision. In the presence of misalignments, our results will still hold as long as Raman side-scattering can be neglected, i.e. when the angle between the two pulses is much smaller than 90$^{\circ}$. The k matching conditions are then still satisfied in the presence of small misalignments because  the non-linear medium, a plasma in the case of our simulations, absorbs any additional transverse wavevector component. Thus, momentum is still locally conserved thereby allowing for Raman backscatter processes (interestingly, we note that when using OAM beams, the wavevectors of seed and pump are already locally misaligned). Despite lowering the total interaction time, and possibly the final amplification level, a small angle between the seed and pump will not change the OAM selection rules and the overall physics of stimulated Raman scattering.

We note that our seed laser pulse final intensity, on the order of $10^{17}~\mathrm{W/cm^2}$, and seed laser spot-size, on the order of $1~\mathrm{mm}$, indicate the production and amplification of Petawatt class twisted lasers with OAM. Additional simulations (not shown) revealed the generation and amplification of circularly polarised OAM modes using a scheme similar to that in Fig.~\ref{fig:setup}b. Moreover, simulations also showed that Raman amplification can also operate when in the absence of exact frequency/wavenumber matching between seed and pump as long as the seed is short so that its Fourier components can still satisfy $k$ and $\omega$ matching conditions. 

Finally, we note that our results could be extended to other non-linear optical media with Kerr non-linearities. In a plasma, the coupling between seed and pump is through an electron Langmuir wave, which also ensures frequency, wavenumber, and OAM matching conditions will hold. In other nonlinear optical media, molecular vibrations, for instance, would play the role of the plasma Langmuir wave. We note that the possibility of OAM transfer has been explored in solids~\cite{bib:martinelli_pra_2004}. Similar phenomenology as illustrated in this work could also be obtained in three wave mixing processes, where an idler wave could play the role of the plasma Langmuir wave. One advantage of testing these setups in non-linear Kerr optical media such as a crystal is that lasers with much lower intensities could be used (see Supplementary note 5 for a discussion in nonlinear optical media with Kerr nonlinearity admitting three-wave interaction processes). The plasma, however, offers the possibility to amplify these lasers to very high intensities. This scheme could also be used in combination with optical pulse chirped pulse amplification (OPCPA) to pre-generate and pre-amplify new OAM modes via stimulated Raman scattering before they enter the plasma to be further amplified. Similar configurations (e.g. stimulated Brillouin backscattering~\cite{bib:alves_subm_2014}) can also be envisaged to produce intense OAM light.

\section{\textbf{Methods}}

\subsection{Setup of numerical simulations and simulation parameters}

Simulations have been performed using the massively parallel, fully relativistic, electro-magnetic particle-in-cell (PIC) code Osiris~\cite{bib:osiris}. In the PIC algorithm, spatial dimensions are discretised by a numerical grid. Electric and magnetic fields are defined in each grid cell and advanced through a finite difference solver for the full set of Maxwell's equations. Each cell contains macro-particles representing an ensemble of real charged particles. Macro-particles are advanced according to the Lorentz force. Since background plasma ion motion is negligible for our conditions, ions have been treated as a positivelly charged immobile background. The plasma was initialised at the front of the simulation box that moves at the speed of light $c$. 
Note that although the simulation are performed in a frame that moves at c, the moving window corresponds to a Galilean transformation of coordinates where all computations are still performed in the laboratory frame.
The simulation box dimensions were $50~\mathrm{\mu m} \times 2870~\mathrm{\mu m} \times 2870~\mathrm{\mu m}$, it has been divided into $650\times2400\times2400$ cells, and each cell contains $1 \times 1 \times 1$ particles ($3.7\times 10^{9}$ simulation particles in total). Additional simulations with $1\times 2\times 2$ particles per cell showed no influence on our conclusions and simulation results. The pump laser was injected backwards from the leading edge of the moving window~\cite{bib:mardahl_mw,bib:mardahl_2}. In order to conserve canonical momentum, the momentum of each plasma electron macro-particle has been set to match the normalised laser vector potential. The particles are initialised with no thermal spread.

The initial OAM seed and pump laser electric field is given by:
\begin{eqnarray}
\label{eq:laguerre}
\mathbf{E} & = & \frac{1}{2}\frac{\mathbf{E_0} w_0}{w(z)} \left(\frac{r\sqrt{2}}{w(z)}\right)^{|\ell|} L_p^{|l|}\left(-\frac{2 r^2}{w^2(z)}\right) \exp\left(-{\frac{r^2}{w^2(z)}}\right) \nonumber \\
& \times & \exp\left[i k \left(z-z_0\right) + \frac{i k z}{1+z^2/z_R^2}\frac{r^2}{z_R^2}  - i (2 p + |\ell| + 1 )\arctan\left(\frac{z}{Z_r}\right)+i \theta_0 + i \ell \phi\right] + c.c.,  
\end{eqnarray}
where $c.c.$ denotes complex conjugate, $\mathbf{E_0}=(E_{0x},E_{0y})$ is the laser electric field at the focus, with $(E_{0x},E_{0y})$ being the electric field amplitudes in the transverse x and y directions respectively. For a linearly polarised laser, there is no phase difference between $E_{0x}$ and $E_{0y}$. For circularly polarised light both components are $\pi/2$ out of phase, i.e. $E_{0x} = \pm i E_{0y}$.
In addition $w^2(z)=w_0^2\left(1+z^2/Z_r^2\right)$ is the waist of the beam as a function of the propagation distance $z$ in vacuum, $w_0$ the waist at the focal plane, $Z_r=\pi w_0^2/\lambda$ is the Rayleigh length, $\lambda=2\pi c/\omega=2\pi/k$ the central wavelength of the laser, $\omega$ and $k$ its central frequency and wavenumber respectively. In addition, $L_{p}^{|l|}$ is a generalised Laguerre polynomial with order $(p,\ell)$, with $\ell$ being the index that gives rise to the orbital angular momentum, $r=\sqrt{x^2+y^2}$ the radial distance to the axis, $\theta_0$ an initial phase, and $z_0$ the center of the laser. We note that all simulations involving Laguerre-Gaussian modes have $p=0$.%
 
The initial electric field of an Hermite-Gaussian (TEM) laser is given by:
\begin{eqnarray}
\label{eq:hermite}
\mathbf{E} & = &\frac{1}{2}\frac{\mathbf{E_0}w_0}{w(z)} H_{m}\left(x\right)H_{n}\left(y\right) \exp{\left(-\frac{r^2}{w^2(z)}\right)} \nonumber \\
& \times &  \exp\left[i k \left(z-z_0\right) + i \frac{k z}{1+z^2/z_R^2}\frac{r^2}{z_R^2} - i \left(m+n\right)\arctan\left(\frac{z}{Z_r}\right) + i \theta_0 \right] + c.c.,
\end{eqnarray}
where $H_m$ is an Hermite polynomial of order $m$. Moreover, the wave-number of the pump laser (which travels in the plasma) in all simulations presented in Figs.~\ref{fig:amp-rate}, \ref{fig:new-oam} and \ref{fig:tem-oam} is set according to the linear plasma dispersion relation $k^2 c^2= \omega^2 - \omega_{\mathrm{p}}^2$, where $\omega_{\mathrm{p}}=\sqrt{4\pi n_0 e^2/m_e}$ is the plasma frequency associated with a background plasma density $n_0$, and where $e$ and $m_e$ are, respectively, the elementary charge and electron mass. The seed frequency and wavenumber are set according to the matching conditions for Raman amplification (see Table 1).

\begin{table}
\centering
\begin{tabular}{l ccccccc}
\hline \hline 
~ & \multicolumn{2}{c}{\textbf{Amplification of}} & \multicolumn{4}{c}{\multirow{2}{*}{\textbf{Generation and amplification of new OAM modes}}} \\ [-8pt]
~ & \multicolumn{2}{c}{\textbf{existing modes}} & \multicolumn{4}{c}{~} \\
\hline 
~ & \multicolumn{2}{c}{~} & \multicolumn{2}{c}{\textbf{OAM seed}} & \multicolumn{2}{c}{\textbf{TEM seed}} \\
~ & \textbf{Pump} & \textbf{Seed} & \textbf{Pump} & \textbf{Seed} & \textbf{Pump} & \textbf{Seed} \\
\hline
\textbf{TEM} & - & - & - & - & \multicolumn{1}{|c}{$\mathrm{TEM}_{00}\mathbf{e}_{\mathrm{x}} + \mathrm{TEM}_{00}\mathbf{e}_{\mathrm{y}}$} & $\mathrm{TEM}_{01}\mathbf{e}_{\mathrm{x}}+\mathrm{TEM}_{10} \mathbf{e}_{\mathrm{y}}$ \\
\textbf{OAM} &	$\mathrm{L}_{00}\mathbf{e}_{\mathrm{x}}$ & $\mathrm{L}_{01}\mathbf{e}_{\mathrm{x}}$ & $\mathrm{L}_{00}\mathbf{e}_{\mathrm{x}}+\mathrm{L}_{02} \mathbf{e}_{\mathrm{y}}$ & $\mathrm{L}_{01}\mathbf{e}_{\mathrm{x}}$ & \multicolumn{1}{|c}{-} & - \\
$\mathbf{a_0~(peak)}$ & 0.02 & 0.06 & 0.02 & 0.03 & \multicolumn{1}{|c}{0.02} & 0.08 \\
$\mathbf{spot~(\mu m)}$ & 718 & 435 & 718 & 435 & \multicolumn{1}{|c}{718} & 435 \\
$\mathbf{duration~(fs)}$ & $25\times10^3$ & 25 & $25\times 10^3$ & 25 & \multicolumn{1}{|c}{$25\times 10^3$} & 25 \\
$\boldsymbol{\omega_0/\omega_{\mathrm{p}}}$ & 20 & 19 & 20 & 19 & \multicolumn{1}{|c}{20} & 19 \\
\hline \hline
\end{tabular}
\label{fig:table}
\caption{\textbf{Laser parameters for the different Raman Amplification regimes to generate and amplify OAM lasers}.Simulation parameters are close to ideal Raman amplification regimes determined by \cite{bib:trines_nphys_2011}. In all simulations, the probe is has a central wavelength of $1~\mu \mathrm{m}$. The background plasma density is $n_0=4.3\times 10^{18}\mathrm{cm-3}$ for all simulations presented. When the pump/probe initially has components in the both transverse directions, the initial peak a0, spot-size and durations present in the table are identical for every component. $\mathrm{L}_n^{\ell}$ refers to a Laguerre-Gaussian $(n,\ell)$ mode, where the OAM corresponds to the index $\ell$. $\mathrm{TEM}_{mn}$ correspond to Hermite-Gaussian lasers with order $(m,n)$. The table only describes the initial simulation conditions.}
\end{table}

\section{\textbf{Acknowledgments}}
Work supported by the European Research Council through the Accelerates ERC project (contract ERC-2010-AdG-267841), by FCT, Portugal (contract EXPL/FIZ- PLA/0834/1012) and by the EU (EUPRAXIA grant agreement 653782). We acknowledge PRACE for access to resources on SuperMUC (Leibniz Research Center).

\section{\textbf{Author contributions}}
All authors contributed to all aspects of this work.

\end{document}